\documentclass[a4paper,10.5pt]{article}
\usepackage{pos}

\usepackage[capitalize]{cleveref}

\let\oldbibliography\thebibliography
\renewcommand{\thebibliography}[1]{%
  \oldbibliography{#1}%
  \setlength{\itemsep}{3pt}%
}

\title{Probing hadronic interactions with measurements at ultra-high energies with the Pierre Auger Observatory}
\ShortTitle{Probing hadronic interactions\ldots}

\manuallySeparateAuthors
\author*[a]{David Schmidt}
\author[b,\dag]{for the Pierre Auger Collaboration}
\notes{\note{Full author list available at: \texttt{http://www.auger.org/archive/authors\_2020\_08.html}}}

\affiliation[a]{Karlsruhe Institute of Technology,\\
Institute of Experimental Particle Physics,\\
Hermann-von-Helmholtz-Platz 1,\\
76344 Eggenstein-Leopoldshafen,\\
Germany}

\affiliation[b]{Observatorio Pierre Auger,\\
Av. San Martín Norte 304,\\
5613 Malargüe,\\
Argentina}

\emailAdd{auger\_spokespersons@fnal.gov}

\renewcommand{\printHeadAuthors}{D. Schmidt}

\abstract{
The characteristics of an extensive air shower derive from both the mass of the primary ultra-high-energy cosmic ray that seeds its development and the properties of the hadronic interactions that feed it.
With its hybrid detector design, the Pierre Auger Observatory measures both the longitudinal development of showers in the atmosphere and the lateral distribution of particles arriving at the ground, from which a number of parameters are calculated and compared with predictions from current hadronic interaction models tuned to LHC data.
At present, a tension exists concerning the production of muons, in that the measured abundance exceeds all predictions.
This discrepancy, measured up to center-of-mass energies of $\sim$ 140 TeV, is irresolvable through mass composition arguments, constrained by measurements of the depth of the electromagnetic-shower maximum.
Here, we discuss a compilation of hadronically-sensitive shower observables and their comparisons with model predictions and conclude with a brief discussion of what measurements with the new detectors of the AugerPrime upgrade will bring to the table.}

\FullConference{%
  40th International Conference on High Energy Physics - ICHEP2020\\
  July 28 - August 6, 2020\\
  Prague, Czech Republic (virtual meeting)
}

\begin{document}
\maketitle

\section{Introduction}
Ultra-high-energy cosmic rays provide access to hadronic interactions at energies well beyond those achievable with human-made accelerators, albeit only through the measurement and interpretation of the extensive air showers that they induce.
These air showers are initiated in collisions of primary cosmic rays with nuclei comprising the upper atmosphere.
What results is a cascade of billions of particles with a footprint in excess of $10$\,km$^2$ at the ground for primaries of the highest energies, which can exceed $10^{20}$\,eV.
Chains of reactions of secondary mesons and baryons form the hadronic cascade of a shower.
Upon losing sufficient energy, the probability for charged mesons to decay exceeds that of further interaction and muons are produced.
These muons serve as tracers of the hadronic component.
The electromagnetic component of air showers is born and fed predominantly from the decay of neutral pions and rapidly proliferates through subsequent cycles of pair production and bremsstrahlung.

Consisting of a hybrid design of fluorescence telescopes overlooking a massive surface detector array,
the Pierre Auger Observatory \cite{ThePierreAuger:2015rma} is the largest detector of extensive air showers in the world.
The fluorescence detector consists of 27 telescopes spread between 4 sites that measure the evolution of extensive air showers as they traverse the atmosphere towards the ground.
It provides a calorimetric measurement of the shower energy through integration of the measured longitudinal profiles of fluorescence light, which is emitted by nitrogen molecules following their excitation by passing charged particles overwhelmingly belonging to the electromagnetic component.
The fluorescence detector additionally measures the depth of the shower maximum $X_\mathrm{max}$, which is a well-studied, mass-sensitive variable that agrees reasonably well in its absolute value and evolution with energy between models.
As the depth of shower maximum inversely scales with mass, showers induced by primaries of lower mass are statistically distinguishable from those of heavier mass.
Overlooked by fluorescence detector is the massive 3000\,km$^2$ surface detector array.
Consisting of over 1600 water-Cherenkov detectors arranged on an isometric triangular grid with 1500\,m spacing, this array samples the lateral distribution of the particles of air showers that arrive at the ground.
From the signal and timing information of each surface detector station, the shower geometry and the energy estimator $S(1000)$, i.e. the signal at 1000\,m from the shower axis, are reconstructed.

A deficit in the number of muons predicted by hadronic interaction models was first observed 20 years ago, and its existence for energies greater than approximately $10^{16}$\,eV has now been corroborated by a number of experiments as summarized in \cite{Cazon:2020zhx}.
With its various hybrid capabilities and unequaled exposure at the highest energies, the Pierre Auger Observatory is equipped to measure a number of hadronically-sensitive air shower observables and has played a key role in confirming and characterizing this deficit.
A compilation of these observables is given here, where it is evident that the measured number of muons exceeds model expectations to an extent irreconcilable by mass composition arguments constrained by complimentary measurements of the electromagnetic component of showers.

\section{Hybrid measurements}

\subsection{Top-down}
\begin{figure}
\centering
\includegraphics[height=0.31\textwidth]{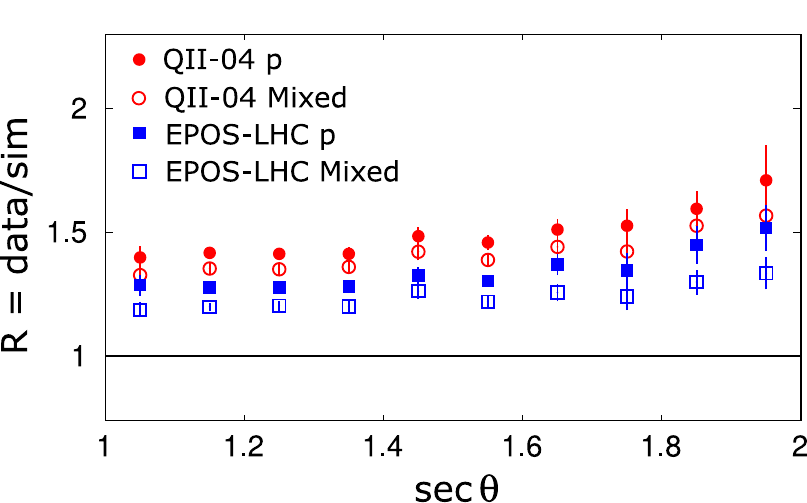}
\hfill
\includegraphics[height=0.3\textwidth]{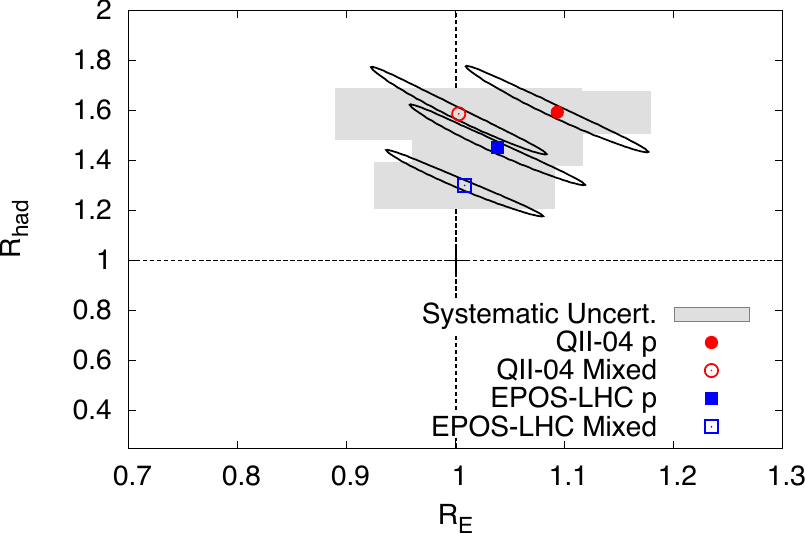}
\caption{
\textit{Left}: Ratio of measured to predicted signals at 1000\,m from the shower axis for simulations best reproducing the measured longitudinal shower profile.
\textit{Right}: The scaling factor for the simulated hadronic signal $R_\mathrm{had}$ and the energy rescaling $R_E$.
Results are shown for two hadronic interaction models and for the $X_\mathrm{max}$ based ``mixed'' and pure-proton assumptions on the mass composition.
The gray rectangles indicate systematic uncertainties.
Figures from \cite{Aab:2016hkv}.
}\label{f:rescalings}
\end{figure}

The signals measured in the water-Cherenkov detectors of the surface detector array derive from both muons and the electrons and photons of the electromagnetic component of showers.
It is not straightforward to decouple the contribution of each component.
For a set of 411 high-quality events measured by both the surface and fluorescence detectors with energies $10^{18.8} <$ $E$/eV $< 10^{19.2}$ (corresponding to 110 to 170 TeV in the center of mass system), however, a novel so-called "top-down" analysis was performed.
For each event, a best-match for the longitudinal profile measured by the fluorescence detector is selected from  batches of shower simulations performed with the primary energy reconstructed by the fluorescence detector.
With a matching simulated shower in hand, the response of the surface detector stations to the particles reaching the ground may be simulated.
The resulting comparison between predicted and measured signals at 1000\,m from the shower axis is shown in \cref{f:rescalings}-left, where it is evident that the measured signal exceeds the predicted one when assuming the mass composition ``mix'' which reproduces the measured distributions of $X_\mathrm{max}$.
This discrepancy holds for the predictions by the different hadronic interaction models, QGSJetII-04 \cite{Ostapchenko:2010vb} and EPOS-LHC \cite{Pierog:2013ria} as well as when assuming a pure proton composition.
An increased discrepancy at larger zenith angles, where the fraction of signal derived from muons is larger, is also visible.

Two scaling factors may be defined.
$R_E$ allows for a shift in the fluorescence detector energy scale, whereas $R_\mathrm{had}$ pertains to signal deriving from the hadronic component of the showers.
Although primarily from muons, this also includes products of muon decay and a small jet component arising from deeply penetrating hadrons.
Since the electromagnetic component of showers is more strongly attenuated in the atmosphere than the muonic and the traversed atmospheric depth increases with zenith angle, both scaling factors can be separately determined given a sizable set of events covering a range of zenith angles.
Values of $R_E$ and $R_\mathrm{had}$ that maximize the likelihood of the signals observed at the ground are obtained and shown in \cref{f:rescalings} (right).
The result of the fit indicates that no energy rescaling is necessary.
The hadronic signal in measurements, however, significantly exceeds that predicted by the models.
Under the assumption of a mixed composition, hadronic rescaling of $R_\mathrm{had}=1.33\pm0.16$ for EPOS-LHC and $R_\mathrm{had}=1.61\pm0.21$ for QGSJet II-04 are required.
Full details of the analysis are covered in \cite{Aab:2016hkv}.

\subsection{Highly Inclined Showers}
\begin{figure}
\centering
\includegraphics[height=0.4\textwidth]{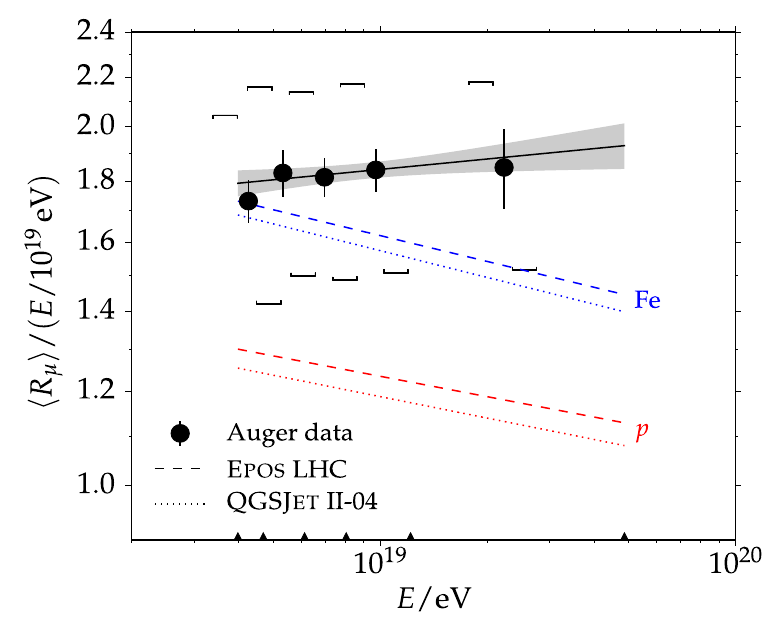}
\hfill
\includegraphics[height=0.42\textwidth]{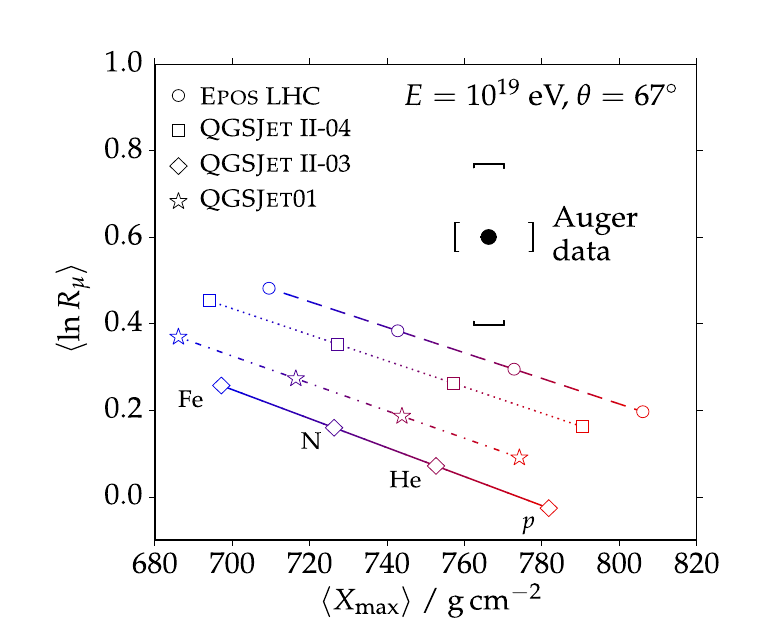}
\caption{
\textit{Left}: Measurements of the average relative muon number $R_\mu$ normalized by the primary energy for inclined showers as a function of energy.
Predictions from two hadronic interaction models for proton and iron primaries are also shown.
\textit{Right}: Juxtaposition of the Auger measurements and model predictions of the average relative muon number and the depth of shower maximum $X_\mathrm{max}$.
Predictions from four hadronic interaction models and four primaries are shown.
Figures from \cite{Aab:2014pza}.
}\label{f:rmu_inclined_showers}
\end{figure}

For zenith angles $\theta > 60^\circ$, the electromagnetic component of air showers is heavily attenuated and the signal in surface detector stations largely derives from muons, facilitating a direct measurement of the muon content.
The lateral distribution for such showers is quite asymmetric due to deflections in the geomagnetic field and for geometric reasons.
The muon density $\rho_\mu$ is modeled at each location in the shower plane $\vec{r}$ as $\rho_\mu(\vec{r}) = N_{19}\,\rho_{\mu,19}(\vec{r};\theta,\phi)$, where $\rho_{\mu,19}$ is a parameterization of the simulated muon density of air showers induced by protons with energies of $10^{19}$\,eV.
Scaling of the density profile of this particular energy by a factor $N_{19}$ is made possible by the fact that the muon density profile has only a very weak dependence on energy and mass and differs little between hadronic interaction models.
Minimizing the bias due to the small deviations from a universal scaling of the profile shape and dependence of the reference profile on the chosen hadronic model results in a parameter $R_\mu$.
The relative muon content $R_\mu$ is shown as a function of the energy reconstructed by the fluorescence detector in \cref{f:rmu_inclined_showers} (left).
The estimated mass composition is evidently at odds with the $X_\mathrm{max}$-based interpretation, as depicted in \cref{f:rmu_inclined_showers} (right).
Through the composition interpretation based on $X_\mathrm{max}$ measurements, the measured number of muons clearly exceeds all model predictions.
The full details of the analysis may be found in \cite{Aab:2014pza}.

\section{Muon Production Depth}
\begin{figure}
\centering
\includegraphics[width=0.95\textwidth]{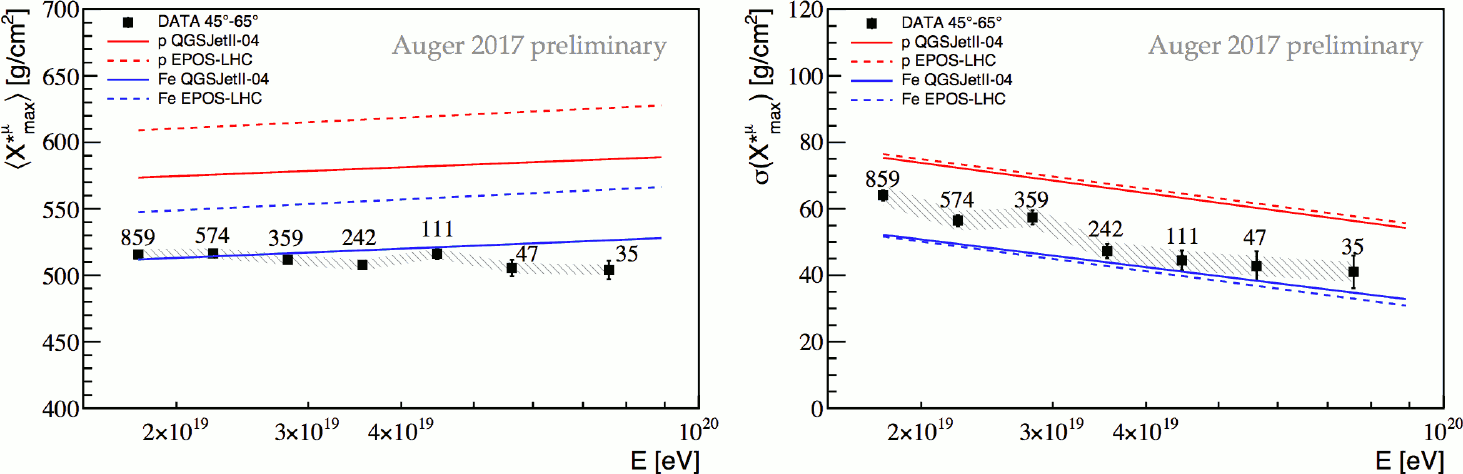}
\caption{
\textit{Left}: Average maximum of the muon production depth distribution.
\textit{Right:} Fluctuations in the maximum of the muon production depth distribution.
Figures from \cite{Mallamaci:2017ovs}.
}\label{f:mpd_xmax_moments}
\end{figure}

At distances $r>1200$\,m, the distance $z$ to the production point of a muon along the shower axis may be calculated from
$z = \frac{1}{2}\left(\frac{r^2}{c\,t_g}-c\,t_g\right)+\Delta-\left<z_\pi\right>$,
where $t_g$ is the time delay of muons with respect to a shower front traveling at the speed of light, $\Delta$ is the distance between the surface detector station and the position of the shower plane at the time of measurement, and $\left<z_\pi\right>$ is the average decay length of the parent meson.
The time delay of muons takes into account the first order geometric delay as well as the parameterized kinematic delay due to sub-luminal muon velocities.
Through knowledge of the atmospheric density profile, the production distance can be converted into a production depth, and the muon production depth distribution can be constructed for all muons for which the described procedure may be applied.
A fit of the muon production depth distribution with a Gaisser-Hillas function provides the depth at which the production rate for observable muons reaching the surface is at a maximum $X^\mu_\mathrm{max}$, which is a parameter defined by both hadronic interactions and mass composition.
\cref{f:mpd_xmax_moments} shows the evolution of both the mean value and fluctuations in $X^\mu_\mathrm{max}$ alongside model predictions for proton and iron primaries.
As with the relative muon content estimator for inclined showers $R_\mu$ described in the previous section, the estimated composition is considerably heavier than that derived from the interpretation using the electromagnetic estimator $X_\mathrm{max}$.
For a full description of the analysis, see \cite{Aab:2014dua,Mallamaci:2017ovs}.

\section{Risetime studies}
The risetime $t_{1/2}$ of the signal in a surface detector station is defined as the duration over which the integrated signal increases from 10\% to 50\% of its total.
Since muons have an earlier average arrival time than the particles of the electromagnetic shower component, an increase in the number of muons results in a smaller value for $t_{1/2}$.

\subsection{Azimuthal asymmetries}
The risetime is largest in the upstream direction of the shower, i.e., where the shower front reaches the ground the earliest and where the traversed atmospheric depth, and therefore the attenuation of the electromagnetic component, is minimal.
The asymmetries of the risetime in the shower plane may be parameterized with
$\left<t_{1/2}\right>/r = a + b\,\mathrm{cos}\zeta + c\,\mathrm{cos}^2\zeta$
where $\zeta$ is the azimuthal angle with $\zeta = 0$ in the upstream direction and where $r$ is the distance from the shower axis.
The asymmetry is non-existent for perfectly vertical showers and initially increases with increasing zenith angle as the difference in atmospheric depth traversed by downstream and upstream stations increases.
At larger zenith angles, however, the asymmetry decreases once more as the electromagnetic component of the showers is more attenuated and the less attenuated muon signal contributes more prominently.
The magnitude of the asymmetry is maximal at the atmospheric depth characterized by $\mathrm{sin}(\theta_\mathrm{max}) $, which in turn depends on the muon content of the showers and thus carries information regarding the hadronic interactions and the primary mass at play.
For a full description of the analysis, see \cite{Aab:2016enk}.


\subsection{Delta method}
\begin{figure}
\centering
\includegraphics[width=0.49\textwidth]{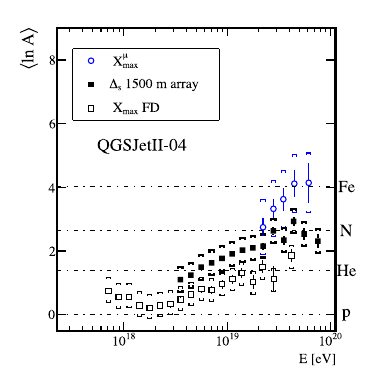}
\hfill
\includegraphics[width=0.49\textwidth]{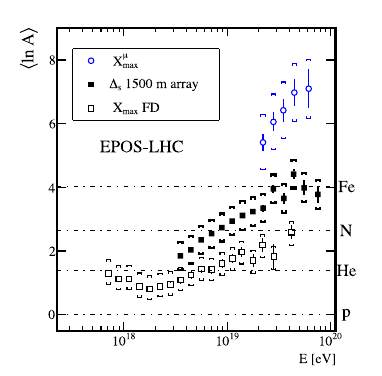}
\caption{
Mean $\mathrm{ln}\,A$ as interpreted using the hadronic interactions models QGSJetII-04 (left) and EPOS-LHC (right) for measurements of the depth of shower maximum $X_\mathrm{max}$, the maximum of the muon production depth distribution $X^\mu_\mathrm{max}$, and with the risetime-based delta method.
Figures from \cite{Aab:2017cgk}.
}\label{f:lna_xmax_delta_xmumax}
\end{figure}

Whereas the previous analysis capitalized on the azimuthal asymmetries in the risetime, the delta method makes use of its dependence on the distance $r$ from the shower axis.
For the stations in an event, the parameter $\Delta_i = (t_{1/2}-t^\mathrm{bench}_{1/2})/\sigma_{t_{1/2}}$ is calculated, where the benchmark $t^\mathrm{bench}_{1/2}$ is defined as the mean behavior of the risetime as a function of\, $r$ observed in the data for a given reference primary energy (hence the value of $\Delta_i$ is, by definition, zero at the reference energy), taking into account zenith-angle dependencies.
The uncertainty $\sigma_{t_{1/2}}$ is also derived from measurements performed with ``twin'' stations separated by only around 10\,m and thereby effectively sampling the same position on the shower plane.
By taking the average $\Delta_i$ of participating stations, $\Delta_s$ is obtained, which evolves linearly in logarithmic energy.
This is analysis is performed considering different distance ranges from the shower axis.
The estimated natural logarithm of the primary mass as interpreted with two hadronic interaction models may be observed in \cref{f:lna_xmax_delta_xmumax} alongside estimations from the depth of shower maximum and the muon production depth.
The apparent mass is compatible with estimations of the risetime asymmetry analysis and can be seen to lie between the the estimates derived from $X_\mathrm{max}$, based on the purely electromagnetic shower component, and that of the muon production depth, based exclusively on the muonic shower component. 
For full details of the analysis, see \cite{Aab:2017cgk}.

\section{Summary and Outlook}
The estimated primary mass based on the depth of shower maximum $X_\mathrm{max}$, an observable derived from the electromagnetic component of showers, is at odds with estimations based on the muonic component.
A compilation of analyses made possible with the hybrid capabilities of the Pierre Auger Observatory were presented here and indicate a deficit in the number of muons predicted by hadronic interaction models.

Resolving the deficit in model predictions of the number of muons requires an increase in the energy routed to the hadronic component of extensive air showers.
Although increasingly disfavored, this could occur in the initial, highest energy interactions, where exotic phenomena could play a role.
Alternatively, this routing could be achieved over many generations of shower development.

The large-scale upgrade to the Pierre Auger Observatory \cite{Aab:2016vlz}, termed AugerPrime, will increase its sensitivity to hadronic interactions and the composition of cosmic rays by disentangling the electromagnetic and muonic components of extensive air showers.
The highlight feature of AugerPrime is a 4\,m$^2$ Scintillator Surface Detector, currently being deployed atop the existing water-Cherenkov detectors.
Its relatively higher sensitivity to the electrons of the electromagnetic shower component will allow for the deconvolution of the muonic and electromagnetic contributions to detector signals.
This disentanglement will improve access to the magnitudes of the different shower components and hadronic interactions on an event-by-event basis.

\end{document}